\documentclass[epj]{webofc}
\usepackage[utf8]{inputenc}
\usepackage[varg]{txfonts}   
\usepackage{booktabs}
\usepackage{xcolor}
\definecolor{darkred}{rgb}{0.4,0.0,0.0}
\definecolor{darkgreen}{rgb}{0.0,0.4,0.0}
\definecolor{darkblue}{rgb}{0.0,0.0,0.4}
\usepackage[bookmarks,linktocpage,colorlinks,
    linkcolor = darkred,
    urlcolor  = darkblue,
    citecolor = darkgreen]{hyperref}
\usepackage{subfigure}
\wocname{EPJ Web of Conferences}
\woctitle{Lattice2017}
\newcommand{\beq}{\begin{equation}}
\newcommand{\eeq}{\end{equation}}
\newcommand{\bea}{\begin{eqnarray}}
\newcommand{\eea}{\end{eqnarray}}
\begin{document}
%
\selectlanguage{english}
\title{%
QCD flux tubes across the deconfinement phase transition
}
\author{%
\firstname{Paolo} \lastname{Cea}\inst{1,2} \and
\firstname{Leonardo} \lastname{Cosmai}\inst{1} \fnsep\thanks{Speaker, \email{leonardo.cosmai@ba.infn.it} } \and
\firstname{Francesca}  \lastname{Cuteri}\inst{3} \and
\firstname{Alessandro}  \lastname{Papa}\inst{4}
}
\institute{%
INFN - Sezione di Bari, I-70126 Bari, Italy
\and
Dipartimento di Fisica dell'Universit\`a di Bari, I-70126 Bari, Italy
\and
Institut f\"ur Theoretische Physik, Goethe Universit\"at, 60438 Frankfurt am Main, Germany
\and
Dipartimento di Fisica, Universit\`a della Calabria, \& INFN - Gruppo Collegato di Cosenza, I-87036 Rende, Italy
}
\abstract{%
We study the behavior across the deconfinement phase transition of the chromoelectric flux tube generated by a static quark and a static antiquark for several distances between them.
We present preliminary results for distances up to 1.33 fm and temperatures up to $1.5 T_c$.}
\maketitle
\section{Introduction}\label{intro}
Lattice formulation of gauge theories allows us to investigate the color confinement phenomenon within a nonperturbative framework. 
Indeed, Monte Carlo simulations produce samples of vacuum configurations that, in principle, 
contain all the relevant information on the nonperturbative sector of QCD.
A wealth  of numerical analyses
in QCD has firmly established that the chromoelectric field between a static quark-antiquark pair
distributes in tubelike structures or ``flux tubes''
~\cite{Fukugita:1983du,Kiskis:1984ru,Flower:1985gs,Wosiek:1987kx,DiGiacomo:1990hc,Cea:1992sd,Matsubara:1993nq,Cea:1994ed,Cea:1995zt,Bali:1994de,Haymaker:2005py,D'Alessandro:2006ug,Cardaci:2010tb,Cea:2012qw,Cea:2014uja,Cea:2014hma,Cardoso:2013lla,Caselle:2014eka,Bicudo:2017uyy,NegroLattice2017}. From these tubelike structures a linear
potential between static color charges naturally arises, thus representing
a numerical evidence of color
confinement.

In our recent studies color flux tubes {\it dominantly}
made up of chromoelectric field directed along the line joining a static 
quark-antiquark pair have been investigated, in the cases of zero temperature~\cite{Cea:2012qw,Cea:2014uja,Cea:2017ocq} 
and nonzero temperature~\cite{Cea:2015wjd}.
In  the present paper we present  new results obtained in studying  the flux tubes across the deconfinement phase transition.
In particular we show preliminary results for several distances between the static quark-antiquark sources and temperatures up to $1.5 T_c$.
The plan of the presentation is as follows: in  section 2 we discuss the observables needed to extract the field strength tensor of the static quark-antiquark sources; in section 3  we  present our numerical results;
and in section 4 our conclusions.
\section{Lattice observables and numerical setup}
\label{background}
\begin{figure}[t] 
\centering
\subfigure[]{\includegraphics[width=0.350\textwidth,clip]{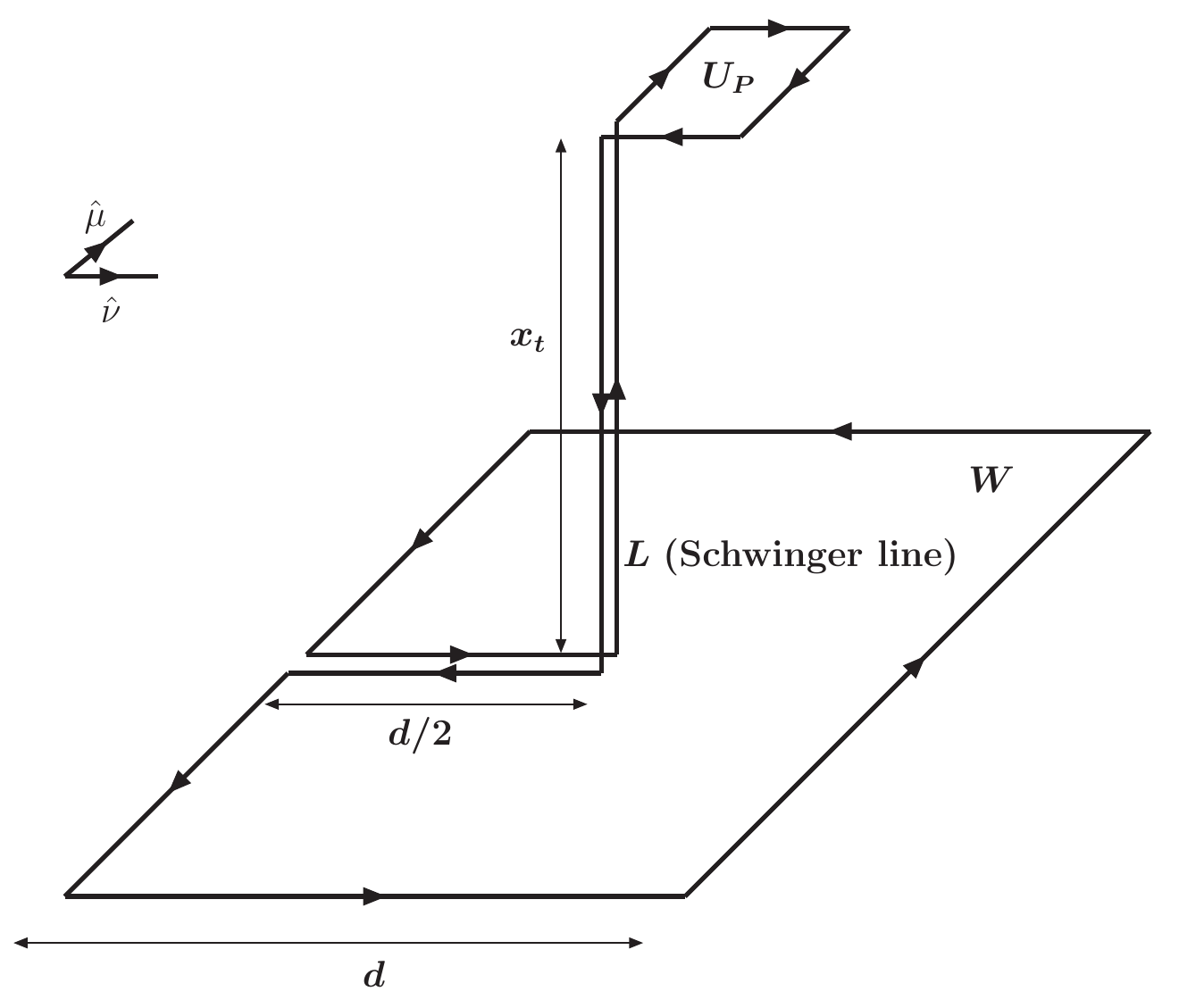}\label{correlators:a}}
\subfigure[]{\includegraphics[width=0.300\textwidth,clip]{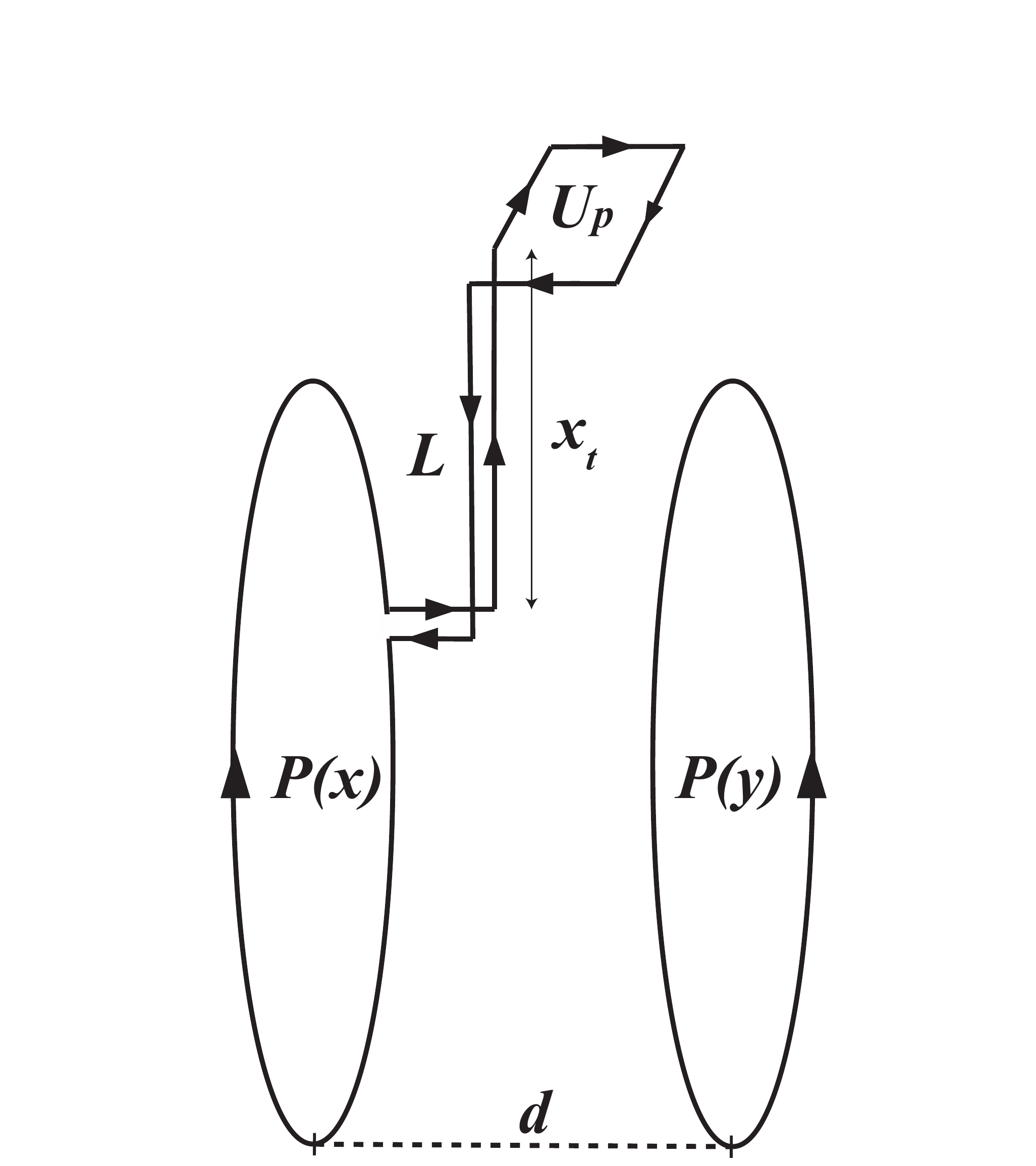}\label{correlators:b}} 
\subfigure[]{\includegraphics[width=0.300\textwidth,clip]{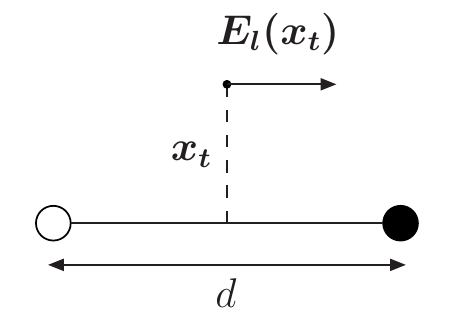}\label{correlators:c}} \\
\caption{(a) The connected correlator given in Eq.~(\protect\ref{rhoW})
between the plaquette $U_{P}$ and the Wilson loop
(subtraction in $\rho_{W,\mu\nu}^{\rm conn}$ not explicitly drawn).
(a) The connected correlator given in Eq.~(\protect\ref{rhoP})
between the plaquette $U_{P}$ and the Polyakov loop
(subtraction in $\rho_{P,\mu\nu}^{\rm conn}$ not explicitly drawn).
(c) The longitudinal chromoelectric field $E_l(x_t)$ with respect to the
position of the static sources (represented by the white and black circles),
for a given value of the transverse distance $x_t$.}
\label{correlators}
\end{figure}

The field distributions generated by a static quark-antiquark pair can be
probed by calculating on the lattice the vacuum expectation value of
the following connected correlation
function~\cite{DiGiacomo:1990hc,Kuzmenko:2000bq}:
\begin{equation}
\label{rhoW}
\rho_{W,\mu\nu}^{\rm conn} = \frac{\left\langle {\rm tr}
\left( W L U_P L^{\dagger} \right)  \right\rangle}
              { \left\langle {\rm tr} (W) \right\rangle }
 - \frac{1}{N} \,
\frac{\left\langle {\rm tr} (U_P) {\rm tr} (W)  \right\rangle}
              { \left\langle {\rm tr} (W) \right\rangle } \; .
\end{equation}
Here $U_P=U_{\mu\nu}(x)$ is the plaquette in the $(\mu,\nu)$ plane, connected
to the Wilson loop $W$, lying on the $\hat 4 \hat i$-plane, with $\hat i$
  any fixed spatial direction, by a Schwinger line $L$, and $N$ is the number
of colors~(see Fig.~\ref{correlators:a}).
In the case of nonzero temperature the role of the Wilson loop is played by two Polyakov loops~(see Fig.~\ref{correlators:b}):
\begin{equation}
\label{rhoP}
\rho_{P,\mu \nu}^{\rm conn}=\frac{\left\langle \mathrm{tr}\left(P\left(x\right)LU_{P}
L^{\dagger}\right)\mathrm{tr}P^\dagger\left(y\right)\right\rangle }{\left\langle 
\mathrm{tr}\left(P\left(x\right)\right)\mathrm{tr}\left(P^\dagger\left(y\right)\right)
\right\rangle }
-\frac{1}{3}\frac{\left\langle \mathrm{tr}\left(P\left(x\right)\right)
\mathrm{tr}\left(P^\dagger\left(y\right)\right)\mathrm{tr}\left(U_{P}\right)\right
\rangle }{\left\langle \mathrm{tr}\left(P\left(x\right)\right)\mathrm{tr}
\left(P^\dagger\left(y\right)\right)\right\rangle}\; ,
\end{equation}
where the two Polyakov lines are separated by a distance $d$.  
The quark-antiquark field strength tensor is obtained as (for a discussion see Ref.~\cite{Cea:2017ocq})
\begin{equation}
\label{fieldstrength}
F_{\mu\nu}(x) = \frac{1}{a^2 g } \; \rho_{W,P,\mu\nu}^{\rm conn}(x)   \; .
\end{equation}
By varying the orientation of the plaquette $U_P$ (Fig.~\ref{correlators:a},~\ref{correlators:b})
it is possible to evaluate all the components of the  chromoelectromagnetic tensor.
We measure the field on the locus of points that are equidistant from the two sources, with $x_t$ measuring the distance between the point of measure and the intersection of the above mentioned locus of points with the $1d$  axis connecting the static sources (Fig.~\ref{correlators:c}).
The numerical results presented here refer to different values of $x_t$ and 
several choices of the distance $d$ between the static sources. 
The role of the distance $d$ between the static sources has been discussed in Ref.~\cite{Baker:2015zlm}.

We performed numerical  simulations  for pure gauge SU(3) 
on $40^3\times 10$  and $48^3\times 12$  lattices, and temperatures
in the range $0.8 T_c \le T \le 1.5 T_c$.
The typical statistics of each run consisted of about 4-5 thousands of configurations;
to allow for thermalization we typically discarded a few thousand sweeps. 
The lattice discretization that we used for the pure gauge SU(3) is the
standard Wilson action, with the physical scale set assuming for the
string tension the standard value of $\sqrt{\sigma} = 420$~MeV and using
the parameterizationgiven in~\cite{Edwards:1998xf}.
For all simulations we made use of the publicly available MILC
code~\cite{MILC}, suitably modified in order to introduce the relevant
observable.
\section{Numerical results}
\label{results}
\subsection{Smoothing procedure}
\label{smoothing}
\begin{figure}[t] 
\centering
\includegraphics[scale=0.4,clip]{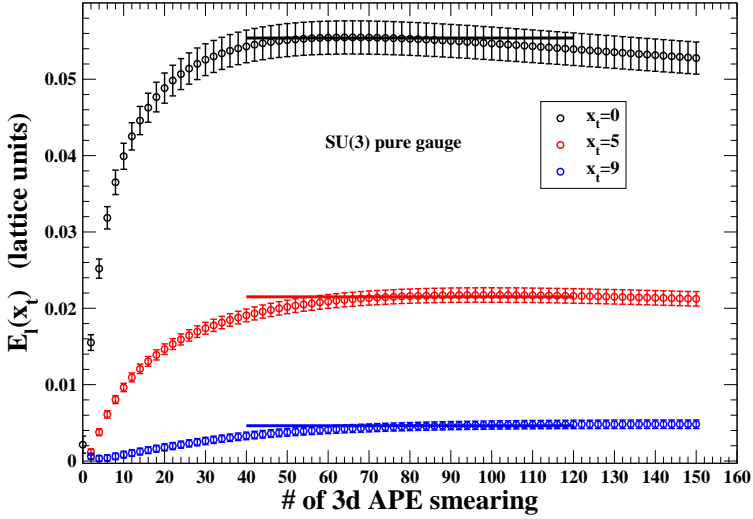}
\caption{Behavior of the longitudinal chromoelectric field
  $E_l$, on a given lattice and for various values of the distance from
  the axis connecting the static sources, {\it versus} the number of
  APE smearing steps on the spatial links.}
\label{El_vs_smear}
\end{figure}
The connected correlator defined in Eq.~(\ref{rhoW}) suffers from large
fluctuations at the scale of the lattice spacing, which are responsible
for a bad signal-to-noise ratio. To extract the physical information carried
by fluctuations at the physical scale (and, therefore, at large distances
in lattice units) we smoothed out configurations by the {\em smearing}
procedure. Our setup consisted of (just) one step of HYP
smearing~\cite{Hasenfratz:2001hp} on the temporal links, with smearing
parameters $(\alpha_1,\alpha_2,\alpha_3) = (1.0, 0.5, 0.5)$, and
$N_{\rm APE}$ steps of APE smearing~\cite{Falcioni1985624} on the spatial links, with 
smearing parameter $\alpha_{\rm APE} = 0.167$. Here $\alpha_{\rm APE}$ is the ratio 
between the weight of one staple and the weight of the original link.
The optimal number of smearing steps was found by looking at the
smearing step at which our direct observable $E_l(x_t)$ showed the
largest signal-to-noise ratio, with the smearing parameter tuned in
such a way that in the $E_l(x_t)$ {\it versus} 'smearing step' plot we could see
a clear plateau.

In Fig.~\ref{El_vs_smear} we show the behavior under smearing of
the longitudinal chromoelectric field $E_l(x_t)$  on a $40^3 \times 10$ lattice at $\beta=6.050$ and
quark-antiquark distance $d=12a$ ($a$ is the lattice spacing) corresponding to a physical distance $d=1.14\,{\textrm{fm}}$.
We can see that, for each value of the distance $x_t$ in the direction transverse
to the axis connecting the sources, a clear plateau is reached after a
sufficiently large number of smearing steps. 
All results concerning the chromoelectric field
$E_l(x_t)$ presented in the following will always refer to determinations
on smeared configurations, after a number of smearing steps $N_{\rm APE}$
such that the plateau is reached for {\em all} considered values of $x_t$.
The typical value of $N_{\rm APE}$ ranges between 50 and 150.
\subsection{Continuum scaling}
\label{continuumscaling}
Our aim is to determine the physical properties of the chromoelectric flux tube
in the {\em continuum}, for this reason, we have preliminarily checked
that our simulations are performed in a region of values of the coupling
$\beta$ where continuum scaling holds.
We have hence measured the longitudinal chromoelectric field generated when
the static sources are located at the {\em same} physical distance $d$, but
for two {\em different} values of the coupling $\beta$, hence at {\it different} distances in lattice units.
\begin{figure}[h] 
\centering
\subfigure[]{\includegraphics[width=0.496\textwidth]{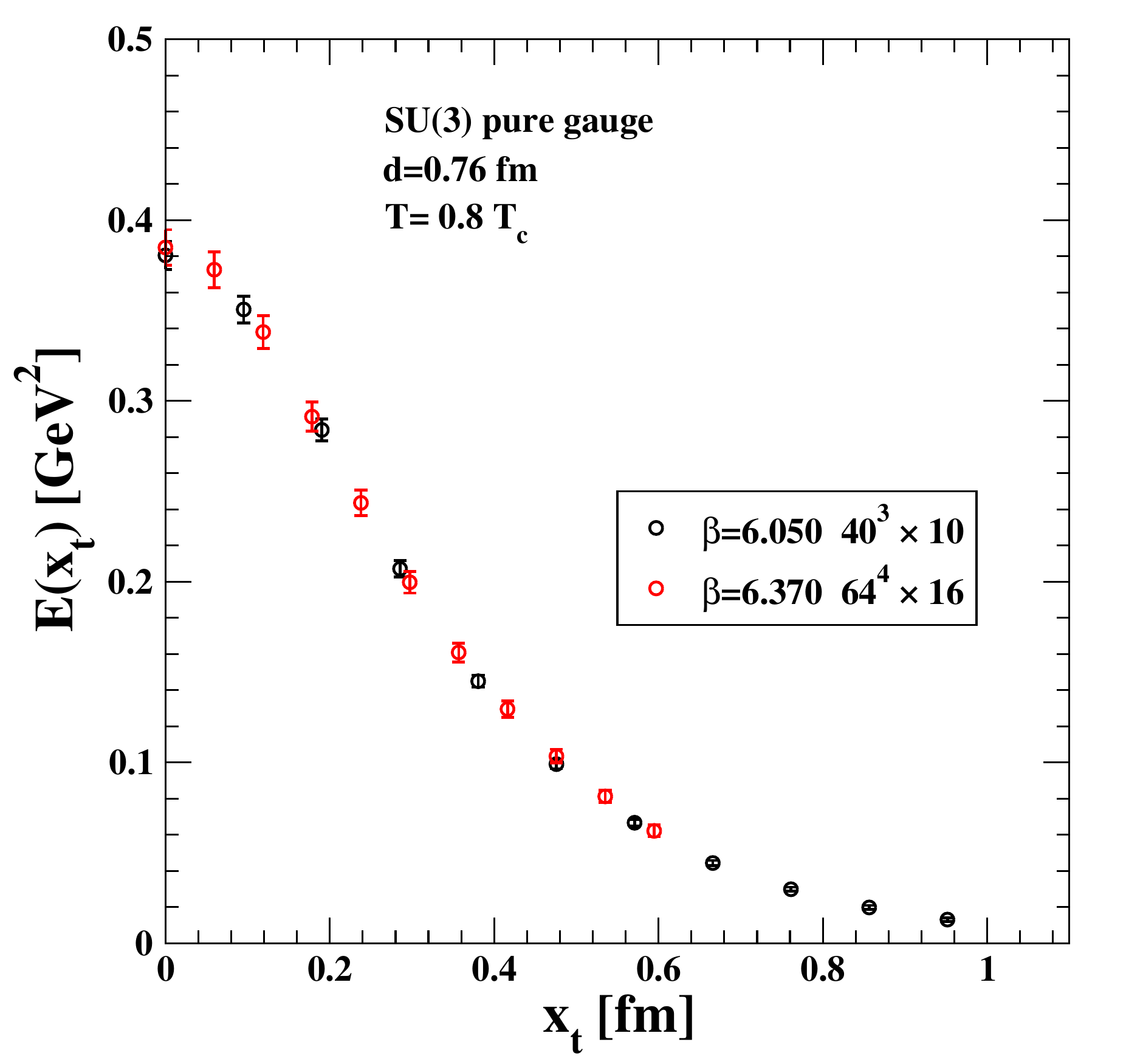}\label{scaling:a}}
\subfigure[]{\includegraphics[width=0.496\textwidth]{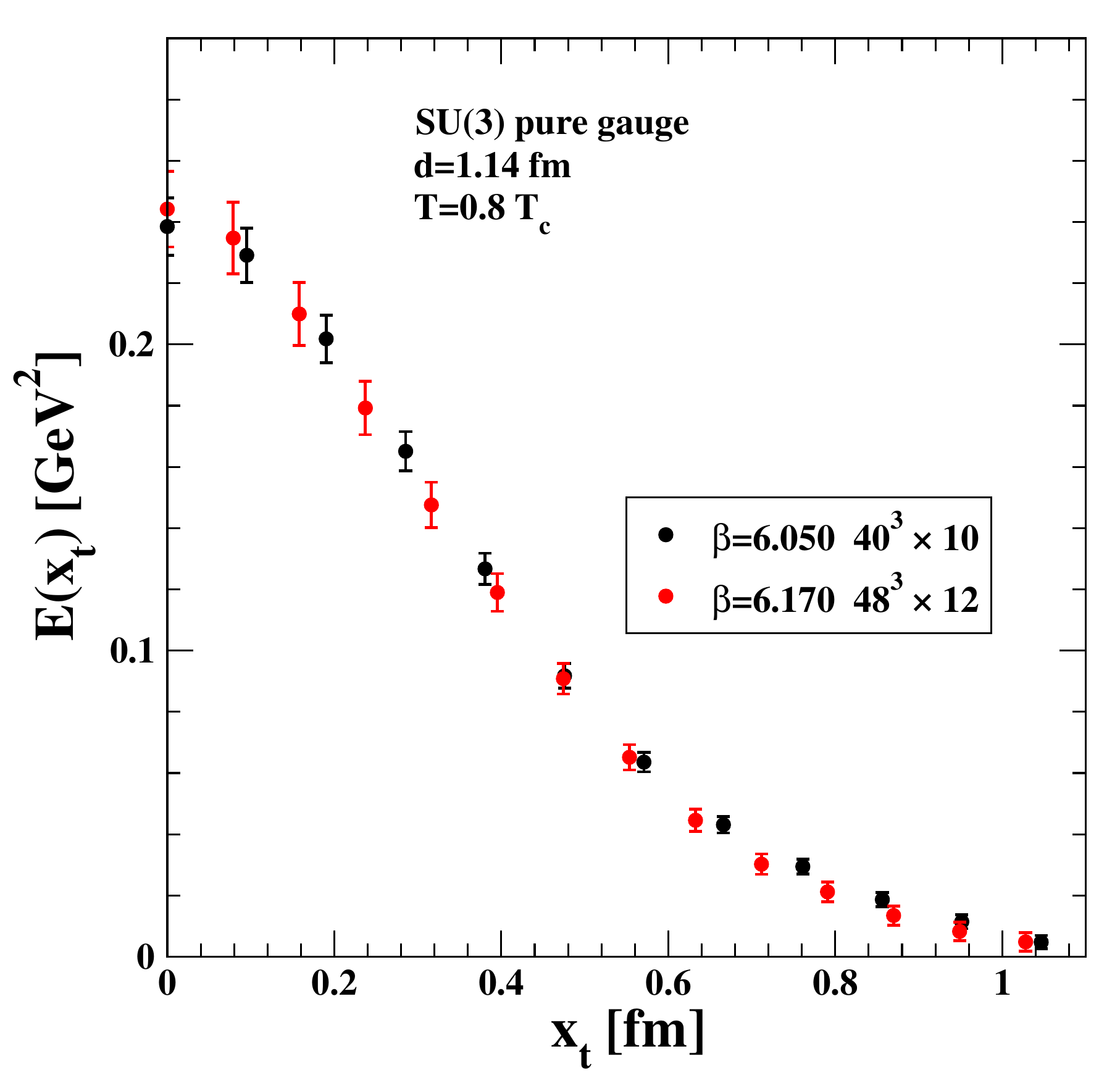}\label{scaling:b}}
\caption{(a) The longitudinal chromoelectric field
  $E_l$ (in physical units) {\it versus} the distance $x_t$ (in physical units)
  from the axis connecting the static sources placed at physical distance $d=0.76\,{\textrm{fm}}$  at $T=0.8T_c$.
  Black open circles refer to $\beta=6.050$ on a $40^3\times 10$ lattice, red open circles refer to $\beta=6.370$ on a $64^3\times 16$ lattice.
(b) The longitudinal chromoelectric field
  $E_l$ (in physical units) {\it versus} the distance $x_t$ (in physical units)
  from the axis connecting the static sources placed at physical distance $d=1.14\,{\textrm{fm}}$  at $T=0.8T_c$. 
Black open circles refer to $\beta=6.050$ on a $40^3\times 10$ lattice, red open circles refer to $\beta=6.170$ on a $48^3\times 12$ lattice.}
\label{scaling}
\end{figure}

In Fig.~\ref{scaling} we present the outcome of this test for two different 
values of the physical distance between the static quark-antiquark pair.
Fig.~\ref{scaling:a} shows  the (smeared) chromoelectric field 
{\it versus} the transverse distance $x_t$ in physical units, when the sources
are placed at distance $8a$ and $10a$  at
$\beta=6.050$ and $\beta=6.195$, respectively, that, in both cases, 
corresponds to a distance roughly equal to 0.76~fm in physical units. 
Fig.~\ref{scaling:b} shows  the (smeared) chromoelectric field 
{\it versus} the transverse distance $x_t$ in physical units, when the sources
are placed at distance $12a$ and $14a$ at
$\beta=6.050$ and $\beta=6.170$, respectively, that, in both cases, 
corresponds to a distance roughly equal to 1.14~fm in physical units.  In both cases an
almost perfect scaling can be observed, thus making us confident that,
for the observable of interest in this work, the continuum scaling is
reached  (at least) for $\beta=6.050$. Another hint from the results shown in 
Fig.~\ref{scaling} is that our use of the smearing procedure is robust: had the
smearing procedure badly corrupted the physical signal for the chromoelectric
field, it would have been quite unlikely to obtain such a nice scaling.
\subsection{Flux tubes across deconfinement}
\label{deconfinement}
\begin{figure}[htb] 
\centering
\includegraphics[scale=0.4,clip]{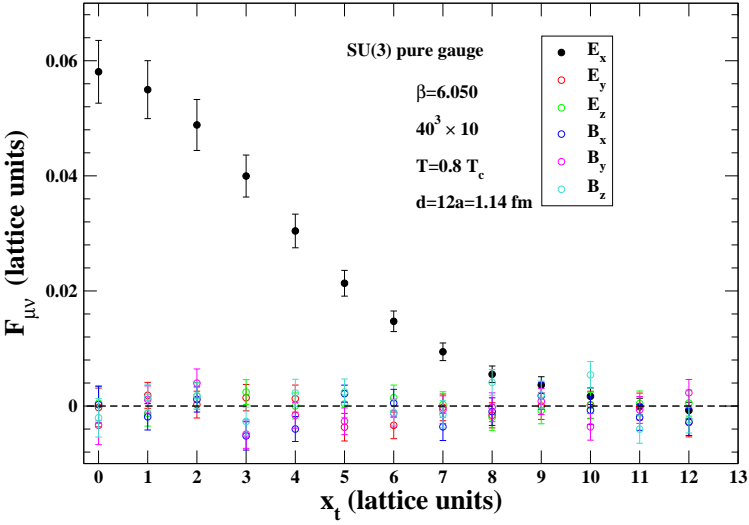}
\caption{The chromoelectromagnetic field (Eq.~(\ref{fieldstrength})) components at $T=0.8T_c$ for quark-antiquark at distance $d=12a=1.14\,{\textrm{fm}}$.}
\label{allfields}
\end{figure}

\begin{figure}[htb] 
\centering
\subfigure[]{\includegraphics[width=0.430\textwidth,clip]{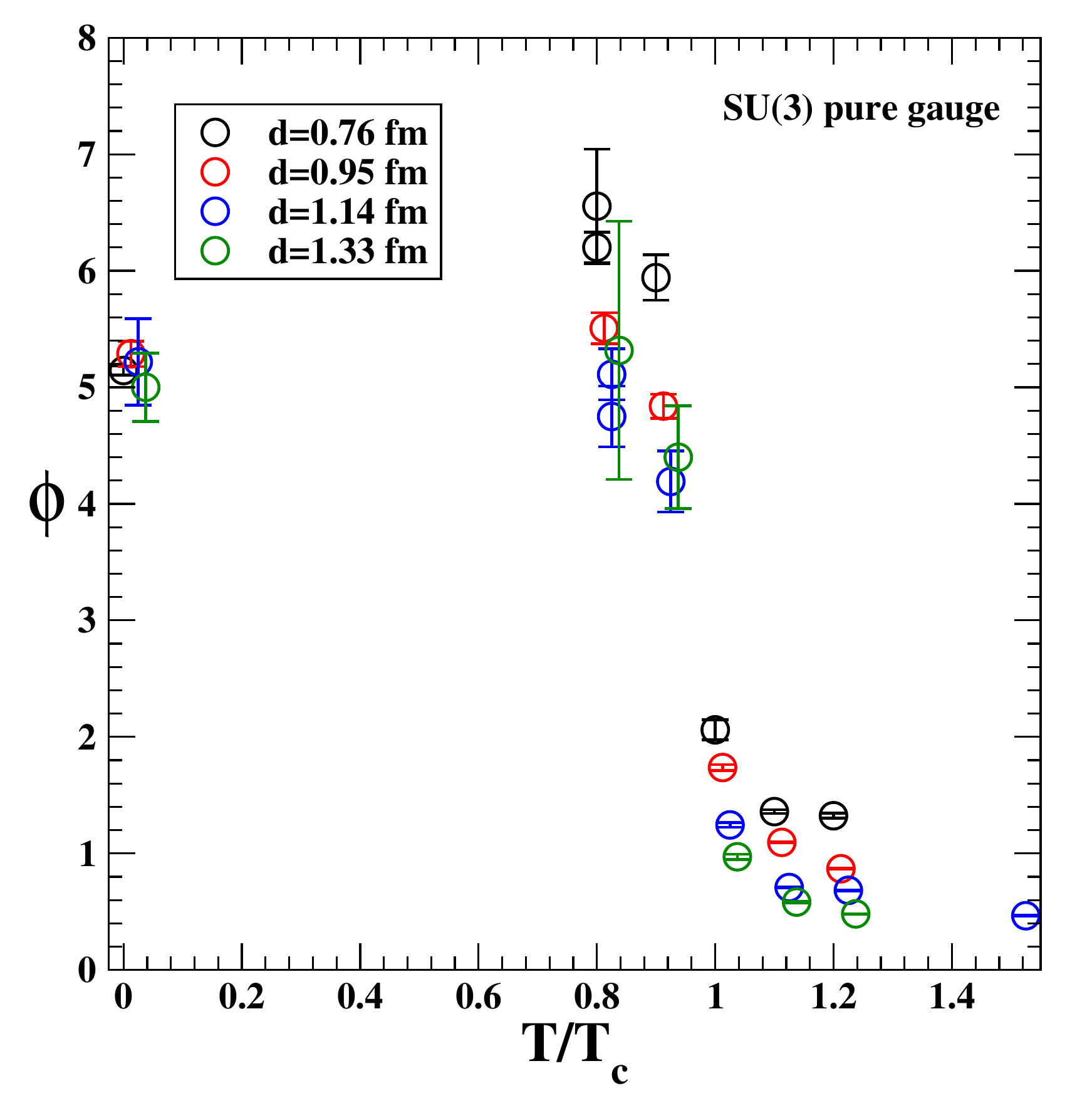}\label{parameters:a}}
\subfigure[]{\includegraphics[width=0.450\textwidth,clip]{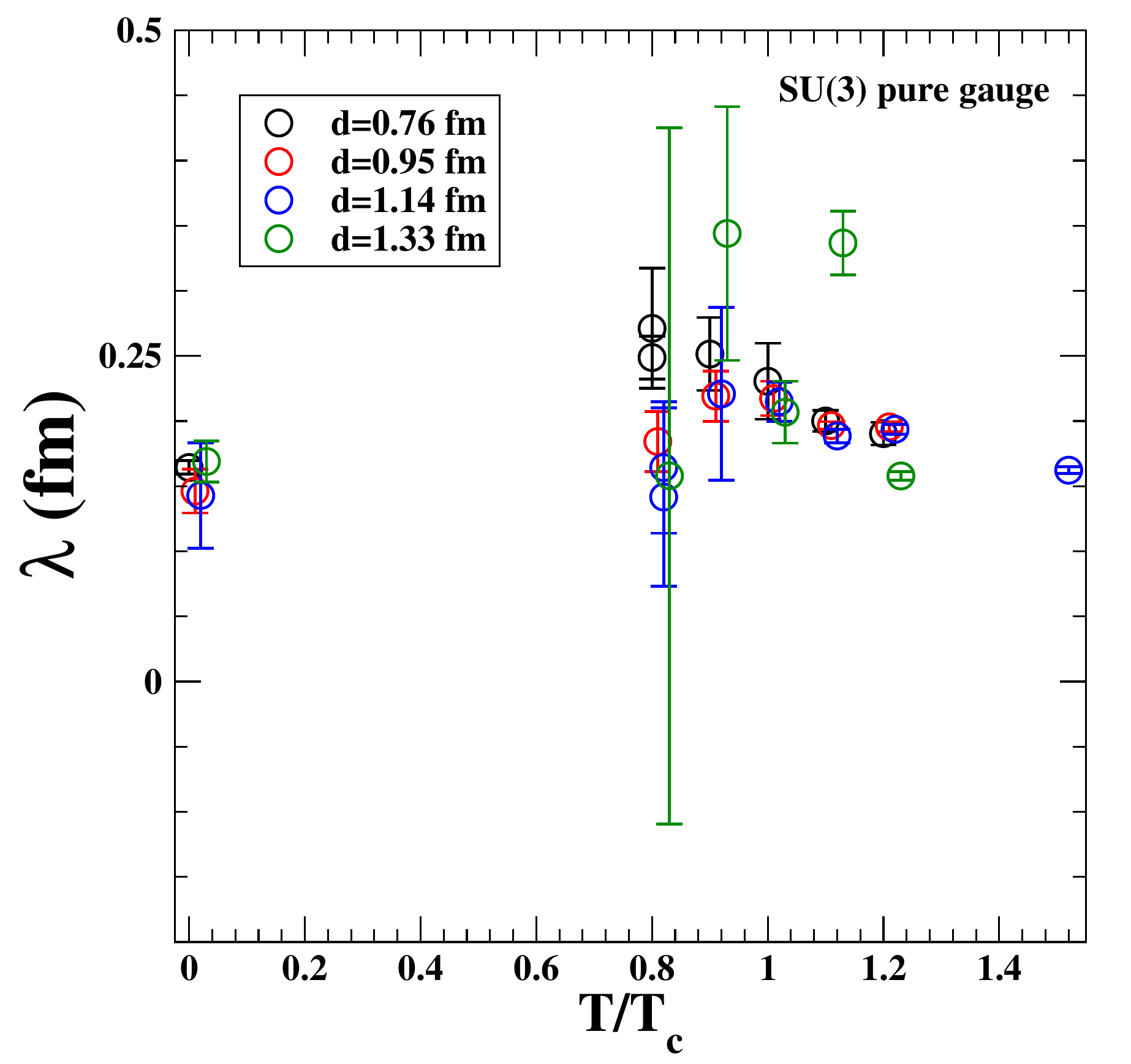}\label{parameters:b}}
\subfigure[]{\includegraphics[width=0.450\textwidth,clip]{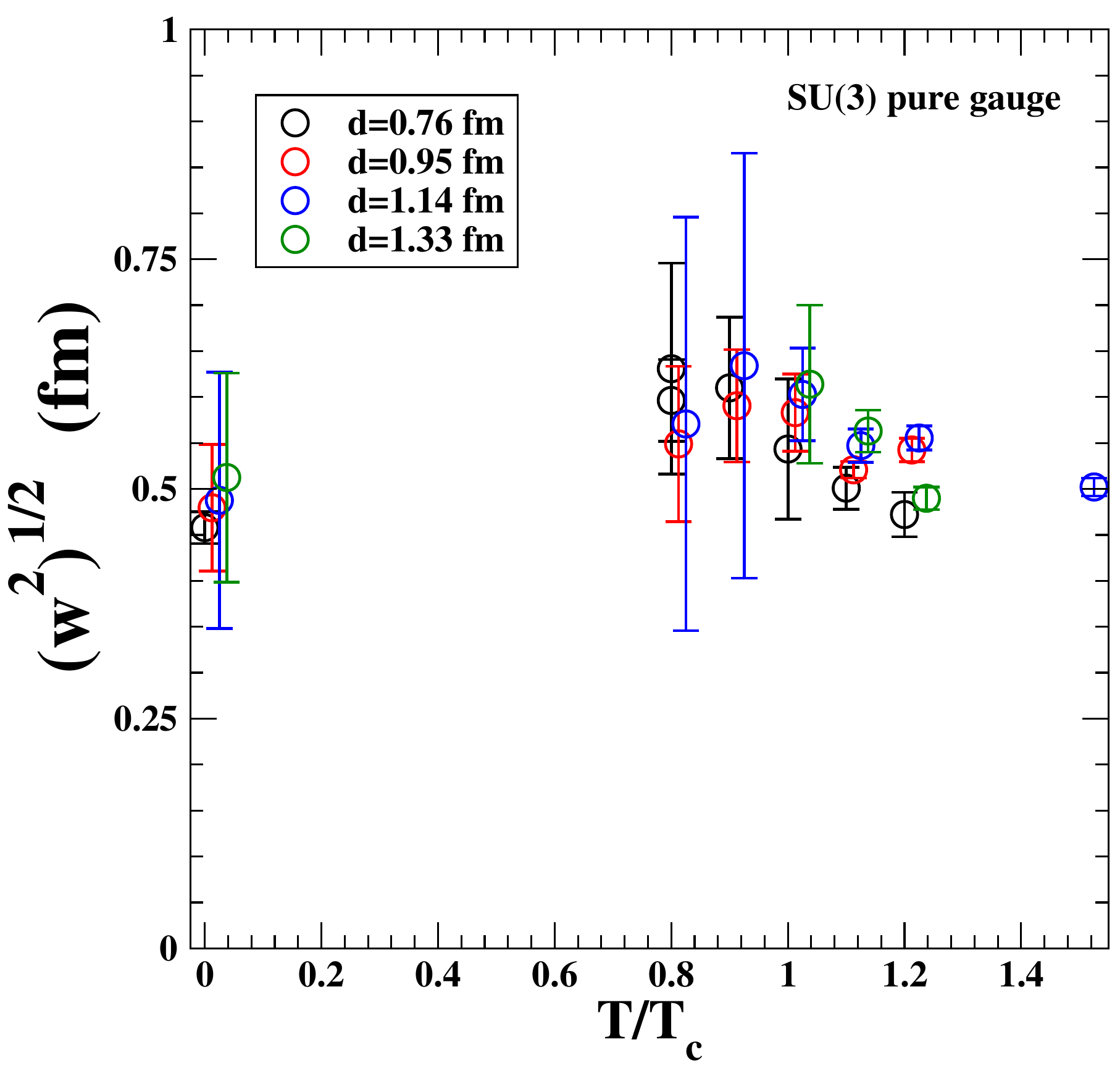}\label{parameters:c}}
\subfigure[]{\includegraphics[width=0.450\textwidth,clip]{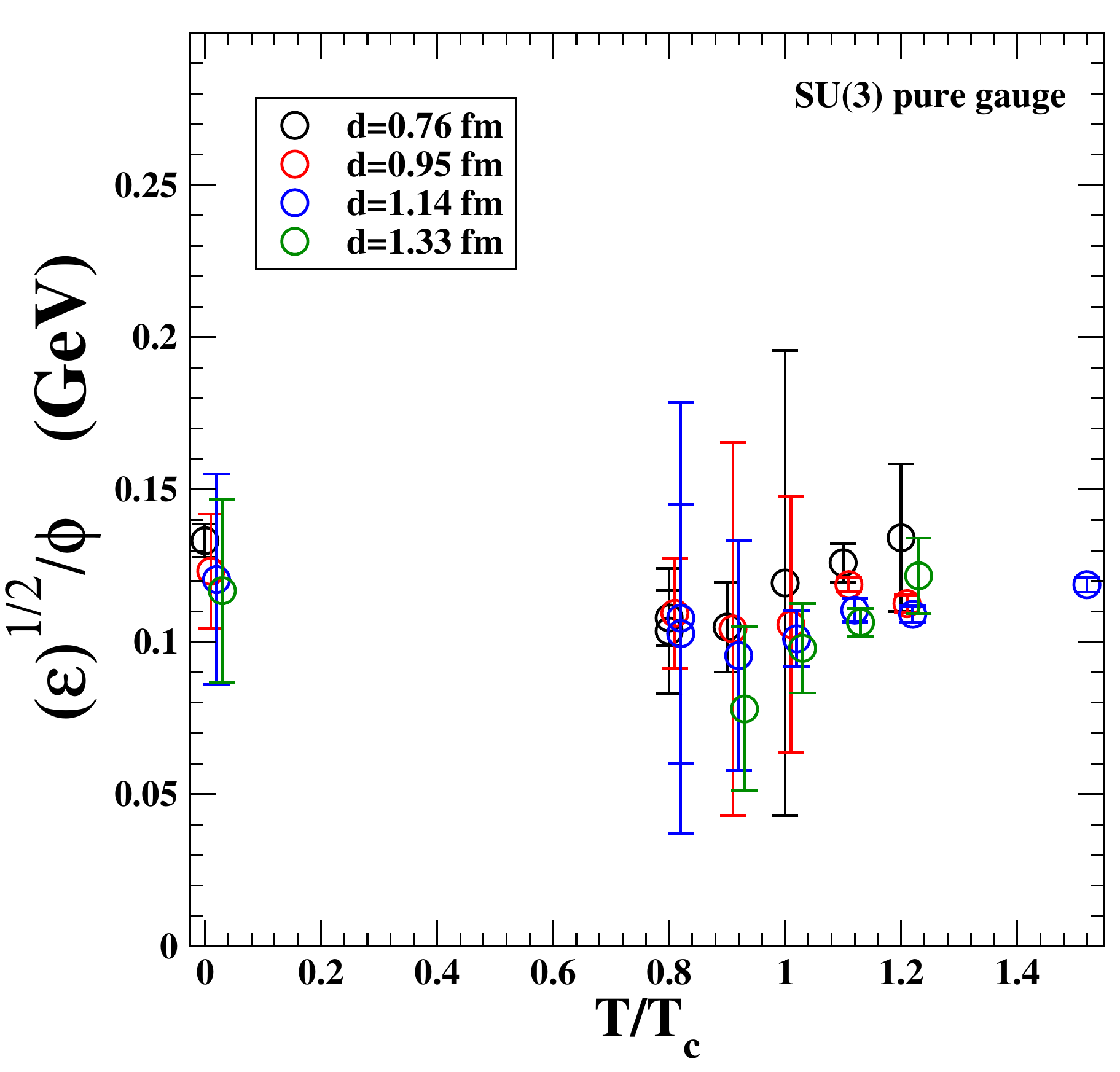}\label{parameters:d}}
\caption{(a) The flux $\phi$ (Eq.~(\ref{Clem})) {\it versus} $T/T_c$ for several values of the distance $d$ between the quark-antiquark sources.
(b) The penetration length $\lambda=1/\mu$ (Eq.~(\ref{Clem})).
(c) The root mean width of the flux tube (Eq.~(\ref{width})).
(d) The energy per unit length in the flux tube (Eq.~(\ref{energy})). Data point abscissas have been slightly shifted for readability.}
\label{parameters}
\end{figure}

We studied the behavior of the flux tubes across deconfinement. 
In particular we measured the transverse shape of the dominant component of the field strength tensor, {\it i.e.} $E_l (x_t )$, having numerically checked, at least for the temperature $T=0.8T_c$, that the other components are zero within statistical uncertainties (see Fig.~\ref{allfields}).
\begin{figure}[t] 
\centering
\subfigure[]{\includegraphics[width=0.450\textwidth,clip]{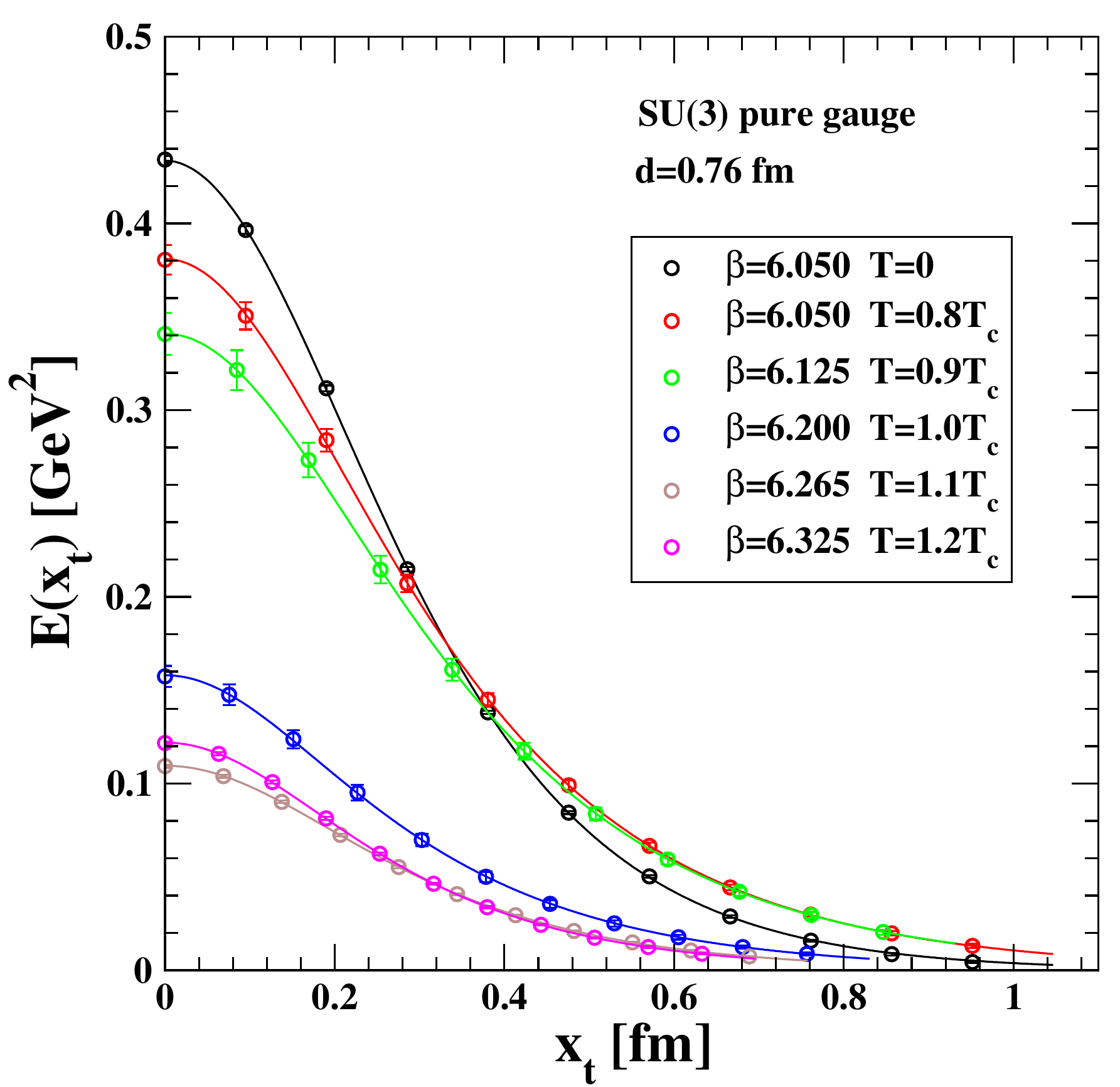}\label{fluxtubes:a}}
\subfigure[]{\includegraphics[width=0.450\textwidth,clip]{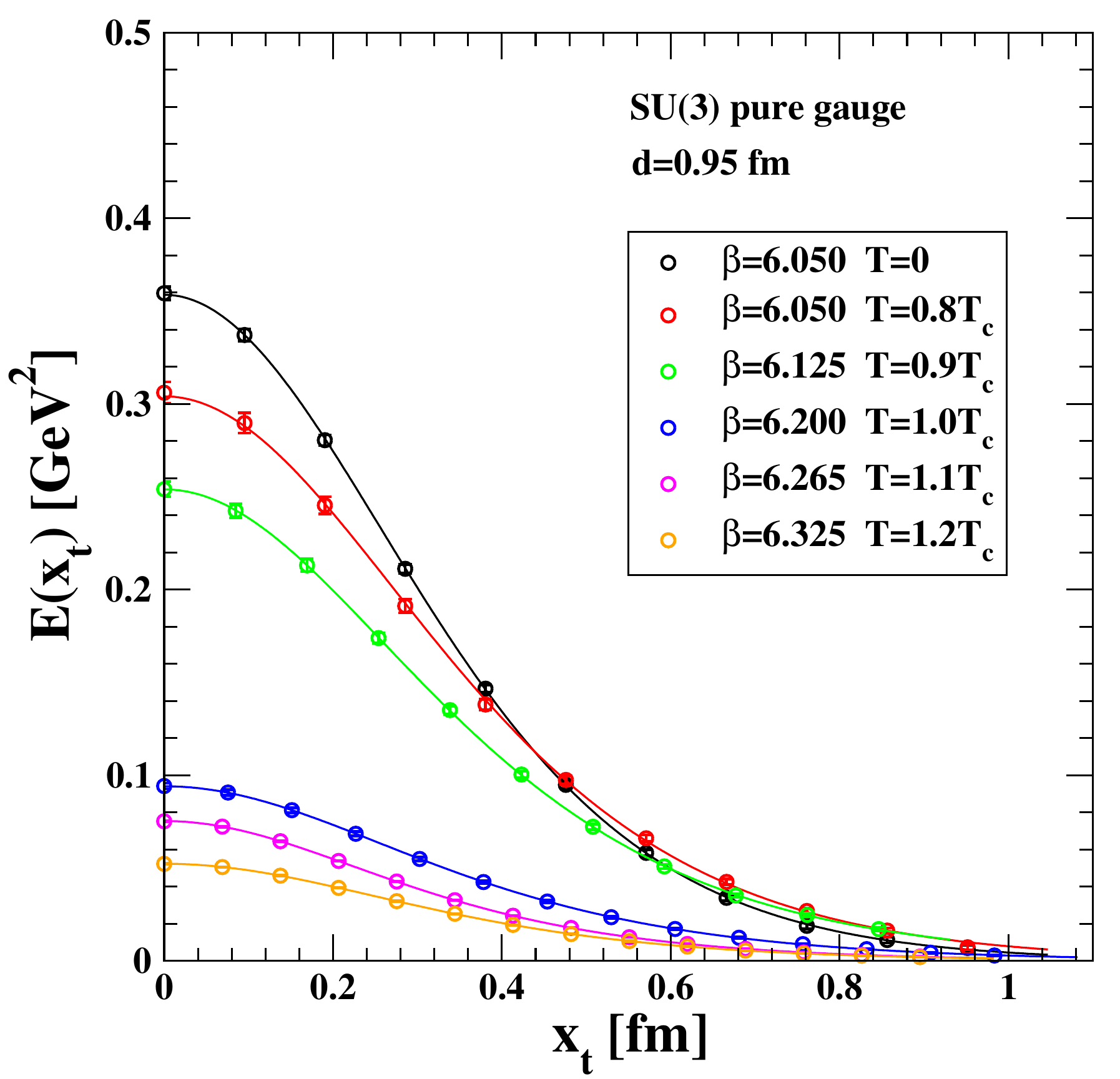}\label{fluxtubes:b}}
\\
\subfigure[]{\includegraphics[width=0.450\textwidth,clip]{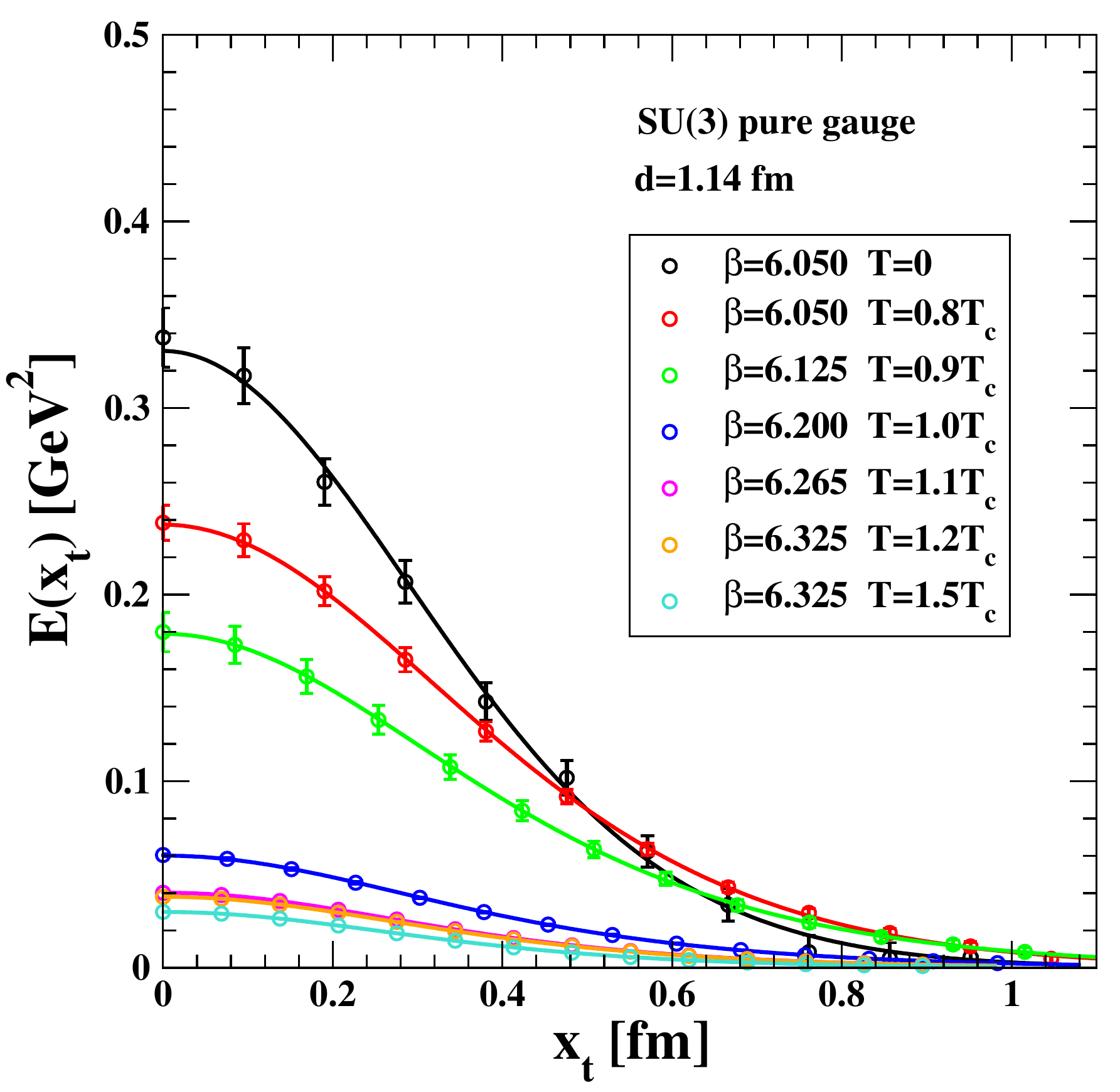}\label{fluxtubes:c}}
\subfigure[]{\includegraphics[width=0.450\textwidth,clip]{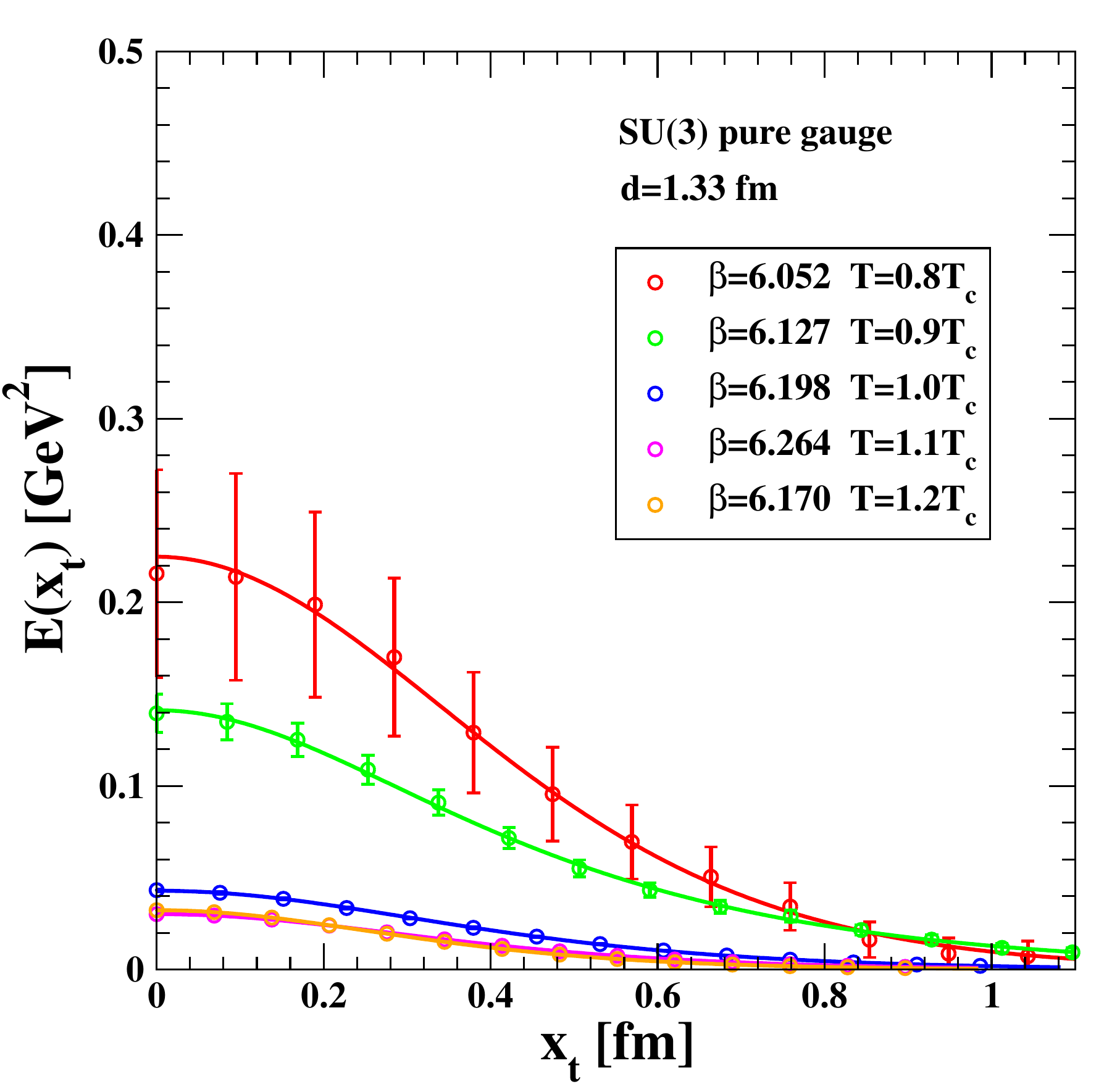}\label{fluxtubes:d}}
\caption{Behavior of the longitudinal chromoelectric field
$E_l$ (in physical units) {\it versus} the distance $x_t$ (in physical units)
from the axis connecting the static sources, at fixed value 
of the physical distance between the sources and  in correspondence of several values of temperature
across deconfinement.}
\label{fluxtubes}
\end{figure}
In Fig.~\ref{fluxtubes} we present the results obtained for the chromoelectric longitudinal field $E_l(x_t)$
in correspondence of temperatures $0.8 T_c \le T \le 1.5 T_c$ with the static quark-antiquark sources
placed apart at distances $0.76\;{\textrm{fm}} \le d \le 1.33\;{\textrm{fm}}$. 
The data for $E_l(x_t)$ are well fitted for all temperatures $T$ and all distances $d$ by using 
a functional form introduced long ago~\cite{Clem:1975aa}  for describing flux tubes in ordinary superconductivity 
and recently proposed~\cite{Cea:2012qw,Cea:2014uja,Cea:2014hma} in order to describe
the transverse distribution of the chromoelectric flux tube:
\begin{equation}
\label{Clem}
E_l(x_t) =  \frac{\phi}{2 \pi} \frac{\mu^2}{\alpha} \frac{K_0[(\mu^2 x_t^2 
+ \alpha^2)^{1/2}]}{K_1[\alpha]} \; .
\end{equation}
In Eq.~(\ref{Clem}) $K_n$ is the modified Bessel function of order $n$, $\phi$ is
the external flux, $\mu=1/\lambda$ with $\lambda$ the London penetration length,
and $\alpha=\xi_v/\lambda$  with $\xi_v$ a variational core radius parameter.
The results in Fig.~\ref{fluxtubes} suggest that the flux tube shape survives across the deconfinement phase transition up to~$T=1.5T_c$,
even though the strength of the field collapses across the phase transition.
By using the fit  Eq.~(\ref{Clem}) to the numerical data for the chromoelectric longitudinal field $E_l(x_t)$
displayed in Fig.~\ref{fluxtubes} it is possible to evaluate the mean square root width of the chromoelectric flux tubes,
\begin{equation}
\label{width}
\sqrt{w^2} = \sqrt{\frac{\int d^2x_t \, x_t^2 E_l(x_t)}{\int d^2x_t \, E_l(x_t)}}
= \sqrt{\frac{2 \alpha}{\mu^2} \frac{K_2(\alpha)}{K_1(\alpha)}} \;,
\end{equation}
and the square root of the {\em energy per unit length}, normalized to the
flux $\phi$,
\begin{equation}
\label{energy}
\frac{\sqrt{\varepsilon}}{\phi} = \frac{1}{\phi} \sqrt{ \int d^2x_t \,
  \frac{E_l^2(x_t)}{2} } =  \sqrt{ \frac{\mu^2}{8 \pi} \,
  \left(1-\left(\frac{K_0(\alpha)}{K_1(\alpha)}\right)^2\right)} \;.
\end{equation}
We can see that the flux $\phi$  drops down across deconfinement (Fig.~\ref{parameters:a}) while 
the penetration length $\lambda$ remains almost constant (Fig.~\ref{parameters:b}), as well as
the mean square root width (Eq.~(\ref{width})) of the flux tube and the square root of the energy per unit length (Eq.~(\ref{energy})).
%
\section{Conclusions}
\label{conclusions}
We presented new preliminary results in studying the flux tube produced by a quark-antiquark pair in the case of SU(3) pure gauge theory across the deconfinement phase transition. 
We have seen that, at least for $T=0.8T_c$, only the chromoelectric longitudinal field contributes to the field inside the flux tube and  the other components are zero within statistical uncertainties.
The shape of the chromoelectric longitudinal field can be well described using a functional form derived from the the ordinary superconductivity~\cite{Clem:1975aa}.
Noticeably the flux tube shape seems to survive across deconfinement up to $T=1.5T_c$, even though the strength of the field collapses across the phase transition.
We plan to check the stability of our results against the change of the smoothing procedure used to get rid of the  large
fluctuations at the scale of the lattice spacing. We also plan to extend our study in pure SU(3) gauge theory to the more realistic case of (2+1)-flavors QCD.
%
\section*{Acknowledgments}
This investigation was in part based on the MILC collaboration's public
lattice gauge theory code. See
{\url{http://physics.utah.edu/~detar/milc.html}}.
Numerical calculations have been made possible through a CINECA-INFN
agreement, providing access to resources on GALILEO and MARCONI at CINECA.


\end{document}